\documentclass{emulateapj}

\newcommand{\zphot}{z_{\mathrm{phot}}}
\newcommand{\zspec}{z_{\mathrm{spec}}}
\newcommand{\dz}{\Delta z}
\newcommand{\dzopz}{\dz / (1+z)}
\newcommand{\dzrange}{\dz < 0.1, 0.2, {\mathrm{and~}} 0.3}
\newcommand{\dzdef}{|\zspec - \zphot|}
\newcommand{\variance}{\sigma^2}
\newcommand{\zvariance}{\sigma_{z}^2}
\newcommand{\rtrain}{r_{\mathrm{train}}}
\newcommand{\rblind}{r_{\mathrm{blind}}}
\newcommand{\rbagging}{r_{\mathrm{bagging}}}
\newcommand{\rcrossval}{r_{\mathrm{cross-val}}}

\slugcomment{Submitted to ApJ}
\shorttitle{Quasar Photometric Redshifts}
\shortauthors{Ball et al.}

\begin{document}

\title{Robust Machine Learning Applied to Astronomical Datasets II: Quantifying
  Photometric Redshifts for Quasars Using Instance-Based Learning}
\author{Nicholas M. Ball\altaffilmark{1,2}, Robert J. Brunner\altaffilmark{1,2},
  Adam D. Myers\altaffilmark{1,2}, \\ Natalie E. Strand\altaffilmark{3}, Stacey
  L. Alberts\altaffilmark{1}, David Tcheng\altaffilmark{2}, Xavier
  Llor\`a\altaffilmark{2}}
\altaffiltext{1}{Department of Astronomy, MC-221, University of Illinois, 1002
  W. Green Street, Urbana, IL 61801, USA}
\altaffiltext{2}{National Center for Supercomputing Applications, MC-257, 1205
  W. Clark St, Urbana, IL 61801, USA}
\altaffiltext{3}{Department of Physics, MC-704, University of Illinois, 1110
  W. Green Street, Urbana, IL 61801, USA}
\email{nball@astro.uiuc.edu}

\begin{abstract}
We apply instance-based machine learning in the form of a $k$-nearest neighbor
algorithm to the task of estimating photometric redshifts for 55,746 objects
spectroscopically classified as quasars in the Fifth Data Release of the Sloan
Digital Sky Survey. We compare the results obtained to those from an empirical
color-redshift relation (CZR). In contrast to previously published results using
CZRs, we find that the instance-based photometric redshifts are assigned with no
regions of catastrophic failure. Remaining outliers are simply scattered about
the ideal relation, in a similar manner to the pattern seen in the optical for
normal galaxies at redshifts $z \lesssim 1$. The instance-based algorithm is
trained on a representative sample of the data and pseudo-blind-tested on the
remaining unseen data. The variance between the photometric and spectroscopic
redshifts is $\variance = 0.123 \pm 0.002$ (compared to $\variance = 0.265 \pm
0.006$ for the CZR), and $54.9 \pm 0.7\%$, $73.3 \pm 0.6\%$, and $80.7 \pm
0.3\%$ of the objects are within $\dzrange$ respectively. We also match our
sample to the Second Data Release of the Galaxy Evolution Explorer legacy data
and the resulting 7,642 objects show a further improvement, giving a variance of
$\variance = 0.054 \pm 0.005$, and $70.8 \pm 1.2\%$, $85.8 \pm 1.0\%$, and $90.8
\pm 0.7\%$ of objects within $\dzrange$. We show that the improvement is indeed
due to the extra information provided by GALEX, by training on the same dataset
using purely SDSS photometry, which has a variance of $\variance = 0.090 \pm
0.007$. Each set of results represents a realistic standard for application to
further datasets for which the spectra are representative.
\end{abstract}


\section{Introduction} \label{Sec: Intro}

Photometric redshifts, both from empirical training sets and template SEDs, are
important for the application of objects to the study of cosmology, as they
enable the exploration of large regions of space that are otherwise
inaccessible. This is achieved both in cosmological volume through a higher
number density of objects and in parameter space through finer binning.

After the early work of \citet{baum:photoz}, \citet{koo:photoz}, and
\citet{loh:photoz}, a variety of techniques were developed extensively
\citep{gwyn:photoz, lanzetta:photoz, mobasher:photoz, sawicki:photoz,
  connolly:angcf, wang:photoz, benitez:photoz} on galaxies in the deep, but
narrow, Hubble Deep Field North, \citep[HDF-N;][]{williams:hdfn}. These
different methods were shown to be mutually consistent and relatively accurate
in blind-testing \citep{hogg:photoz}.

More recently, wide-field surveys with multicolor photometry and fiber-based
spectroscopy have generated large, uniform samples that enable photometric
redshifts to be estimated for both galaxies and quasars.

For galaxies in these surveys at redshifts of $z \lesssim 0.4$,
\citep[e.g.,][]{brunner:photoz, brunner:angcf, tagliaferri:nnphotoz,
  firth:annphotoz, vanzella:hdfannphotoz, ball:ann, collister:annz,
  wadadekar:photoz} a number of results have converged to an RMS dispersion of
$\sigma \sim 0.02$ (i.e., $\variance \sim 0.0004$) between spectroscopic and
photometric redshifts, with no serious systematic effects. It should be
emphasized, however, that galaxy photometry in these previous analyses has been
very good, typically a few percent or better. \citet{way:photoz} show similar
results when combining the SDSS DR2 \citep{abazajian:dr2}, GALEX GR1
\citep{martin:galex} and the extended source catalog of the 2 Micron All Sky
Survey \citep{skrutskie:2mass}.

The results at moderate redshifts have also been successful, with luminous red
galaxies \citep{eisenstein:lrgsample} in the SDSS trained with redshifts in the
2SLAQ survey \citep{cannon:2slaqlrg} having an RMS of $\sigma = 0.049$
\citep{collister:megazlrg} for a sample at $0.4 < z < 0.7$ \citep[see
also][]{padmanabhan:photoz}.

At high redshifts, the number of spectra available is smaller and, in addition
to the HDF-N, there have been analyses of other deep fields such as the
HDF-South \citep{williams:hdfs} and the Hubble Ultra Deep Field
\citep{beckwith:hudf}. In the latter, \citet{coe:photoz} show an accuracy of
$\dz = 0.04(1+z)$ for $z \lesssim 6$.

In contrast to galaxies, which show small numbers of outliers but no significant
groups of outlying objects, all wide-field quasar photometric redshift results
to date \citep{richards:qsophotoz, budavari:qsophotoz, weinstein:sdssqsophotoz,
  wu:qsophotoz, babbedge:impz} suffer from regions of `catastrophic' failure, in
which groups of objects are assigned a redshift very different from the true
value. The first four use SDSS data, while the latter uses the ELAIS N1 and N2
fields and the Chandra Deep Field North. \citet[][hereafter
W04]{weinstein:sdssqsophotoz} implement an empirical method based on
color-redshift relations, which we use as our baseline. Catastrophic failures
severely hamper cosmological investigations that use photometrically selected
quasar samples
\citep[e.g.,][]{myers:qsoevoln,myers:nbckdeall,myers:nbckdesmall}, particularly
by assigning objects at $z > 2$ to $z < 1$ and vice-versa, thus eliminating
these regions is important. Reasons for the failures, depending on the details
of the way a particular dataset is chosen, include quasar reddening, degeneracy
in the color-redshift relation, and superimposition of emission from another
object, for example, an extended host galaxy.

Results using a more restricted parameter space \citep{wolf:qsophotoz}, defined
by $17 < R < 24$ and $1.2 < z < 4.8$ in the 17 filter set of the COMBO-17 survey
\citep[e.g.,][]{wolf:combo17}, have met with more success. However the sample
size, 192 quasars, is small, and limited in angular extent, and therefore is of
limited cosmological applicability.

In this paper we utilize optical data from the Fifth Data Release of the SDSS,
and near- and far-UV data from the Second Data Release of the Galaxy Evolution
Explorer \citep[GALEX;][]{martin:galex} to assign photometric redshifts to
quasars. Our results improve upon previous wide-field techniques, by eliminating
regions of catastrophic failure, resulting in a distribution of quasar
photometric redshifts comparable to those obtained for galaxies. We do not
address the application of the photometric redshifts to any parameter space
beyond that represented by the training and blind test sets.

\section{Data} \label{Sec: Data}

We utilize data from the Fifth Data Release (DR5, SDSS collaboration, in
preparation) of the Sloan Digital Sky Survey \citep[SDSS,][]{york:sdss} and the
Second Data Release (GR2) of the Galaxy Evolution Explorer
\citep{martin:galex}. We select primary non-repeat observations of objects
classified as quasars ({\tt specClass} = {\tt qso} or {\tt hiz\_qso}) in the
{\tt specObj} view of the SDSS DR5 Catalog Archive Server database. The {\tt
  hiz\_qso} objects are at redshifts of $z > 2.3$ and trigger the use of the
Lyman $\alpha$ finding code in the SDSS spectroscopic pipelines (Frieman et al.,
Schlegel et al., in preparation). We also require that the spectroscopic flags
{\tt zWarning} = 0 and {\tt zStatus} $>$ 2, and that all input magnitudes are
not at clearly unphysical extreme values, being in the range 0--40. The 
resulting sample contains 55,746 quasars.

In addition to the SDSS sample, the SDSS objects are cross-matched to the
primary photometric objects in the {\tt photoObjAll} view of the GALEX GR2
database. We find 8,174 matches within an RA+DEC tolerance of 4 arcsec. 532 of
these have more than one match and are rejected, leaving an SDSS+GALEX sample of
7,642 unique matches. For the GALEX objects, we require {\tt primary\_flag} = 1,
a detection in both near and far-UV bands, magnitudes again in the range 0--40,
and the flags {\tt fuv\_artifact} and {\tt nuv\_artifact} to be 0.

Throughout, the SDSS magnitudes are corrected for Galactic extinction using the
dust maps of \citet{schlegel:dustmaps} and the GALEX magnitudes using the $B-V$
({\tt e\_bv}) term inferred from these maps using the standard formula of
\citet{cardelli:extinction}.

The resulting samples of 55,746 and 7,642 objects form training sets used as
input for the learning algorithms. The full set of object attributes for the
SDSS sample consists of 16 training features. These are the colors $u-g$, $g-r$,
$r-i$, and $i-z$, where the SDSS bands $u$, $g$, $r$, $i$, and $z$ are given for
each of the four magnitude types PSF, fiber, Petrosian, and model
\citep{stoughton:edr}. For SDSS+GALEX, we add the colors $fuv-nuv$ and $nuv-u$,
where $u$ is given in each of the four SDSS magnitude types, resulting in 21
training features.

In addition to the SDSS and SDSS+GALEX datasets, we also analyze the SDSS+GALEX
sample of objects, but using only SDSS features. This dataset, referred to as
{\it GALEX-SDSS-only}, enables us to quantify the level of improvement in
SDSS+GALEX seen from the addition of the GALEX UV features, as opposed to
possible improvement due to the sample only containing quasars that appear in
both SDSS and GALEX.

\section{Algorithms} \label{Sec: Algorithms}

We implement instance-based learning on the SDSS, SDSS+GALEX and GALEX-SDSS-only
datasets. The results are compared to those on the same data for an empirical
color-redshift relation containing full probability density functions (Strand,
in preparation). We also study the utility of subsets of the full set of
training features using genetic algorithms.

The machine learning is implemented in the Java environment Data-to-Knowledge
\citep{welge:d2k}. It is optimized through use of nationally peer-reviewed
allocated time on the Xeon Linux cluster Tungsten at the National Center for
Supercomputing Applications. This enables an extensive exploration of the
parameter space describing the training features of the objects and the settings
of the learning algorithms.

\subsection{Instance-based Learning} \label{Subsec: IB}

Instance-based learning \citep[IB,
e.g.][]{aha:ib,witten:datamining,hastie:learning}, is a powerful class of
empirical machine learning methods that to date has not been extensively
utilized on large astronomical datasets due to its computational intensity.
Two examples where the method has been used are \citet{budavari:qsophotoz} and
\citet{csabai:edrphotoz}, who both use the method on the SDSS Early Data Release
\citep[EDR][]{stoughton:edr}. However, they only utilize single nearest neighbors,
and in addition the DR5 dataset analyzed herein is approximately 15 times the 
size of the EDR. Here, through the use of Tungsten (\S \ref{Sec: Algorithms}, 
above), we are able to realize the full potential of the algorithm, via the use 
of the $k$-nearest neighbor method \citep[e.g.,][]{cover:nn}.

In its simplest form, the `training' of the algorithm is trivial, and involves
simply memorizing the positions of each of the examples in the training set. For
each object in the testing set, the nearest training example is then found, and
the predicted value, either a classification or a continuous value, is taken to
be that of the training example. Thus the computational expense is incurred at
the time of classification, as a large number of distance calculations must be
performed. However, the method is powerful because it uses all of the
information available in the training set, rather than a model of the training
set as is typically used by most other learning algorithms.

There are a number of simple refinements to this method, which in practice
result in large improvements in performance: (1) Instead of the nearest neighbor
to the testing example, the $k$ nearest neighbors can be found, and the
distances weighted using a predictive integration function to produce a weighted
output. This function, $d$, takes the form $$d = \sum_{i}^{k} \frac{1}{x_i^p},$$ 
where the $x_i$ are the Euclidean distances to the neighbors, and the exponent
$p$ can take on any positive value, typically but not necessarily an integer. (2) 
The input features can be {\it standardized} such that the mean and variance of 
each are 0 and 1 respectively. This stops the training being dominated by 
features with larger numerical values or spreads. Alternatively, one could also 
normalize the range of features to be 0--1. (3) Objects in the training set can 
be allocated to collective regions of parameter space, which can considerably
reduce the required number of distance calculations.

Of the methods described, we implement (1) and (2), but not (3) as we wish to use
the full information available in the training data. We optimize the values of
$k$ and $p$ and standardize all training features. Further refinements can also
be made for objects which have non-continuous values such as a classification or
missing data. However, in this paper all values are considered, i.e., the
training features and the spectroscopic and photometric redshifts, are
continuous.

\subsection{Color-Redshift Relation} \label{Subsec: CZR}

We have implemented the color-redshift relation (CZR) method of
\citet{weinstein:sdssqsophotoz} on the same data as the IB. This enables a
direct comparison of the performance of the two methods. The CZR establishes an
empirical relation between the spectroscopic redshifts and the colors of the
training set. The maximum likelihood redshift Probability Density Function (PDF)
is then found for each object in the test set.

\subsection{Genetic Algorithms} \label{Subsec: GA}

The methods above select and optimize a learning algorithm for a given set of
training features. However, it is possible that different subsets of the
features available will produce better results. In particular, the results for
instance-based learning can be made worse by noise in the training set or by
irrelevant training features. To explore this possibility, we implement a binary
genetic algorithm on the training feature sets.

A genetic algorithm \citep[GA:
e.g.,][]{holland:genetic,goldberg:genetic,haupt:genetic} mimics evolution, in
the sense that the most successful individuals are those that are best adapted
for the task at hand. We implement the binary genetic algorithm, in which each
individual is a string of 0s and 1s which represents whether or not to use a
particular input feature (in our case the 16 colors). An initial population of
random individuals is created and the IB is run using the features selected. The
result, in this case the variance between photometric and spectroscopic
redshift, is the {\it fitness} of that individual. The individuals and their
fitnesses are then combined to produce new individuals, and those with higher
fitnesses are favored. In principle, a good approximation to the best set of
features to use as the training set should be selected with this approach.

The combination involves identifying the best individuals to breed via
tournament selection, in which a specified number of individuals from the
population are selected and the best is put in the mating pool to be combined
with other individuals. Two individuals are combined using one point
crossover, in which a segment of one is swapped with that of the other. To more
fully explore the parameter space and prevent the algorithm from converging too
rapidly on a local minimum, a probability of mutation is introduced on the newly
created individuals before they are processed. This is simply the probability that
 a 0 becomes a 1, or vice-versa.

An approximate number of individuals to use is given by $$n_{in} \sim
2n_f~log(n_f),$$ where $n_f$ is the number of features. Here, for the SDSS and
GALEX-SDSS-only, $n_f=16$, and for SDSS+GALEX, $n_f=21$. Hence, $n_{in} \sim 39,
56$ respectively for these two values of $n_f$. The algorithm converges,
i.e., finds the best individual and hence the best training set, in $$n_{it} \sim
\alpha n_f~log(n_f)$$ iterations, where $\alpha$ is a problem-dependent
constant. Generally $\alpha > 3$, giving an expected value for our data of
$n_{it} \sim 58$ for $n_f=16$, and $n_{it} \sim 83$ for $n_f=21$. We employ this 
number of iterations with larger numbers of individuals\footnote{300 for 
SDSS+GALEX, and 200 for the other two datasets. These numbers were selected for 
other tests not reported here, and simply strengthen the null result.} to be sure 
that the algorithm has converged. Further information on genetic algorithm design 
can be found in, e.g., \citet{goldberg:design}.

Our GA is implemented on the IB for each of the SDSS, SDSS+GALEX and
GALEX-SDSS-only datasets. The settings of these algorithms are fixed for the
duration of the GA iteration. It is possible in principle to combine the
optimization of the learning algorithm and the feature set; however, we defer
this analysis to a later paper.

\subsection{Training and Quality of Redshifts} \label{Subsec: Training}

The IB and CZR are supervised learning algorithms---they are given a training
set of objects and attempt to minimize a cost function which describes the
quality of the predictions on a separate testing set. 

For IB, the cost function is given by the variance between the photometric and
spectroscopic redshifts for objects with spectra: $$\left<(\dz)^2\right> -
\left<\dz\right>^2,$$ where $\dz = \dzdef$, $\zspec$ is the spectroscopic
redshift value, and $\zphot$ is the photometric redshift prediction made by the
learning algorithm. The second term in the variance equation is small. 

The value of the variance is dominated by the outliers. However,
in our case, this is a desirable property, because it is these objects which we
wish to pull in the most toward the correct values. The dominance of the
outliers renders the variance susceptible to variations in this population. We
therefore quote errors on all of our blind test variances, derived from
splitting the population using multiple random seeds (see below). For the CZR,
the cost function is the likelihood of the PDF.

Instance-based learning, like any supervised machine learning algorithm, is
susceptible to incompleteness and noise in the training set. At the present
time, the SDSS DR5 is by far the largest and most homogeneous quasar dataset 
available, and it has a high completeness \citep[e.g.,][]{vandenberk:empirical}.
Other available datasets are either not as deep, smaller 
\citep[e.g.,][]{croom:qsolf}, or deeper but orders of magnitude smaller 
\citep[e.g.,][]{wolf:combo17}. One could prune noisy exemplars, however, it is 
difficult to meaningfully define what is a noisy or sparsely populated region of 
parameter space, and pruning particular regions could introduce new and poorly 
defined biases. The use of multiple nearest neighbors smoothes the noise, and 
the blind test results address both incompleteness and noise by presenting 
realistic results on unseen data.

The distance measure parameters of a number of nearest neighbors and the 
distance weighting assume that the input training features are uncorrelated, 
however, given that we repeat the same four colors in four magnitude types, and 
in addition that a set of features is always derived from a particular object, 
the input features will always be correlated, both in magnitude type (e.g., PSF 
$u-g$ is correlated to fiber $u-g$, and so on), and in color (e.g., PSF $u-g$ is
correlated to PSF $g-r$, and so on.) Correlated input features are therefore
unavoidable; we feel, however, that our algorithmic approach is acceptable 
because we select the parameters to produce the optimal blind test result.

Different splits of the training set are investigated at various points in the
learning process, giving four adjustable ratios: (1) $\rtrain$ is the ratio
between the data used as the training set and for testing the algorithm's
performance according to the cost function to adjust the final model settings
(for IB there is no adjustment so the ratio just affects the performance through
the information available). (2) $\rblind$ is the ratio of the whole set of data
used in training and testing to that unseen by the algorithm until it is
applied, as it would be to new data from another survey; this is the
pseudo-blind test. (3) $\rbagging$ is the ratio of the data used in each bagged
model to the rest of the training data, where the training data is $\rtrain$ of
the whole dataset. (4) $\rcrossval$ is similar, but for cross-validation. The
latter is distinguished from bagging because it takes different random
subsamples of the whole $\rtrain$ training and $1-\rtrain$ testing set, whereas
bagging subsamples $\rtrain$.

The value for which we quote results for all of these ratios is 80:20. For
application to new data not used here, the value of $\rtrain$ would be 100\%,
to maximize the information available. This is the standard $\sigma^2$ reported
in the literature for CZR techniques, but its value would be meaningless for
instance-based approaches.

For IB, the variances obtained are quoted from the pseudo-blind test, as this
represents the most realistic standard of performance available from within the
SDSS and GALEX datasets to be expected on new data. $\rblind$ is always such
that the training data is representative of the full dataset.

We quote the mean and standard deviation of the best variance from ten training
runs with differing random seeds for $\rblind$. Each run produces a grid of
models with the range $1 \leq k \leq 50$ and $1 \leq p \leq 10$, where $k$ is
the number of nearest neighbors and $p$ the exponent in the distance-weighting
function (\S \ref{Subsec: IB}). Integral values of $k$ and $p$ were used,
although this is not a requirement. We use positive values of $p$ as negative
values would result in objects other than the nearest neighbor being given the
highest weighting, which would be unphysical as increasingly large values of $k$
would be given an ever higher weight. We investigated bagging and cross-validation
using values of $\rbagging$ and $\rcrossval$ of 80:20 and 50:50 but these were
not found to be necessary for IB. Other measures, such as $\dzopz$, and the
percentage of objects within $\dzrange$ are also given for comparison to other
work. We do not quote any results in which there is any overlap between the
training and testing data.

The comparative CZR results were obtained by using a 10-fold bootstrapped
pseudo-blind test, again in the ratio $\rblind$ = 80:20.

\section{Results} \label{Sec: Results}

We now describe results for the full SDSS DR5, SDSS DR5 + GALEX GR2, and
GALEX-SDSS-only datasets, all of which are summarized in Table \ref{Table:
  photoz}.

\subsection{SDSS DR5}

\begin{figure}
\figurenum{1}
\plotone{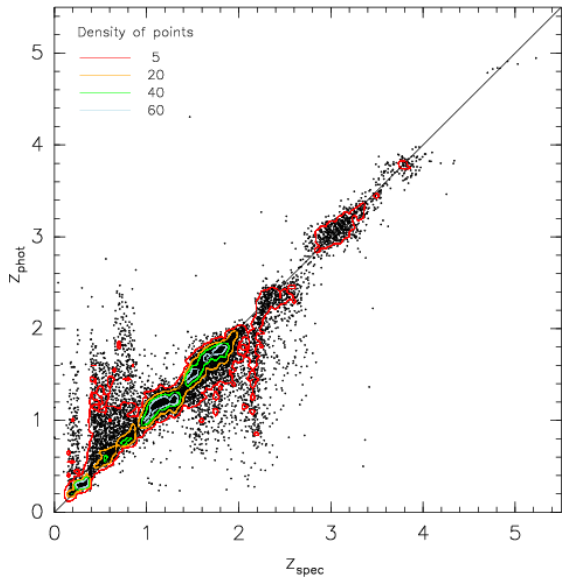}
\caption{Contour plot of quasar photometric redshifts assigned by the
  instance-based learner versus spectroscopic redshifts for the SDSS DR5
  pseudo-blind testing sample of 11,149 of 55,746 quasars described in the
  text. For contouring, the objects are placed in bins of 0.05 in redshift,
  although the values on both axes are continuous. The variance between the two
  measures over the whole redshift range is $0.123 \pm 0.002$. Compared to
  Figure \ref{Fig: CZR photoz}, there are no regions of `catastrophic' failure,
  in which objects are assigned a very different redshift to the true value,
  just a smoothly declining spread of outliers. There are no objects outside the
  range of redshifts plotted. \label{Fig: SDSS photoz}}
\end{figure}

\begin{figure}
\figurenum{2}
\plotone{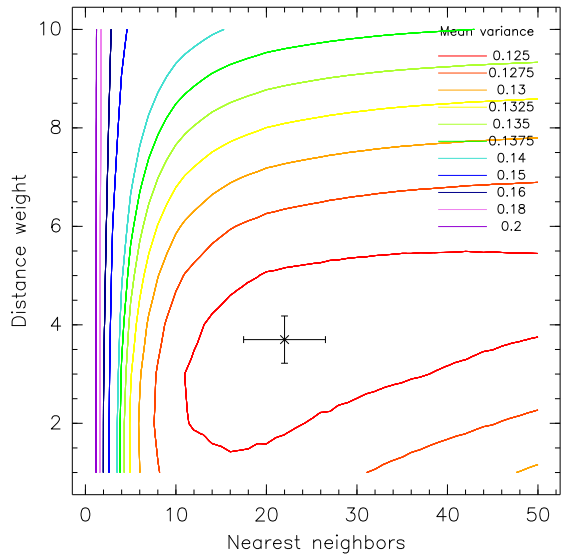}
\caption{Effect of varying the number of nearest neighbors and the distance
  weighting of the instance-based learner for the pseudo-blind test on the SDSS 
  DR5 dataset, showing the mean from ten different training to pseudo-blind test
  splits of the data with a varying random seed. The model which gives the lowest 
  variance is marked with $1\sigma$ error bars. \label{Fig: SDSS grid}}
\end{figure}

We found that the ideal parameters are $22 \pm 5$ nearest neighbors (NN) and a
distance weighting (DW) of $3.7 \pm 0.5$. In the pseudo-blind test on the unseen
20\% of the data, the best variance between the photometric and spectroscopic
redshifts is $0.123 \pm 0.002$. A comparison between the photometric and
spectroscopic redshifts is shown in Figure \ref{Fig: SDSS photoz}, and the
effect of varying the NN and the DW for the pseudo-blind test is shown in Figure
\ref{Fig: SDSS grid}. We find that $54.9 \pm 0.7\%$, $73.3 \pm 0.6\%$, and $80.7
\pm 0.3\%$ of the objects are within $\dzrange$, respectively. The variance
weighted by redshift is $\zvariance = 0.034 \pm 0.001$ and the mean $\dzopz =
0.095 \pm 0.001$.

Because the values of NN and DW used here are discrete (in principle they can be
continuous, but that was not attempted), the results presented in Figure
\ref{Fig: SDSS photoz} were obtained with the values of NN, DW and the blind
test set random seed that gave the best variance in its grid that was closest to
the mean. Here, these values are ${\mathrm{NN}}=22$, ${\mathrm{DW}}=4$ and a
random seed of 8 (for the seeds we used the integers 0 to 9). The variance is
0.1240, which is consistent with the mean variance quoted.

Our key result, shown in Figure \ref{Fig: SDSS photoz} is the absence of regions
of catastrophic failure---there is no upturn in a histogram of $\dz$ values at
large $\dz$, just a smooth decline such that few objects are outliers. This is
in contrast to previous results for quasar photometric redshifts, which, while
showing a comparable spread of objects with low $\dz$, show outlying regions of
objects with high $\dz$. The scattering of outliers obtained by the IB is
similar in form to that seen in other studies for normal galaxies at redshifts
of $z \lesssim 1$ (see, for example, Figure 3 of \citealt{ball:ann} for SDSS
Main Sample galaxies, which have a mean redshift of $z \sim 0.1$), although
there is still structure seen in Figure \ref{Fig: SDSS photoz}, especially at
$\zspec \lesssim 1, {\mathrm{and~}} \zspec \sim 2.2$.

\begin{figure}
\figurenum{3}
\plotone{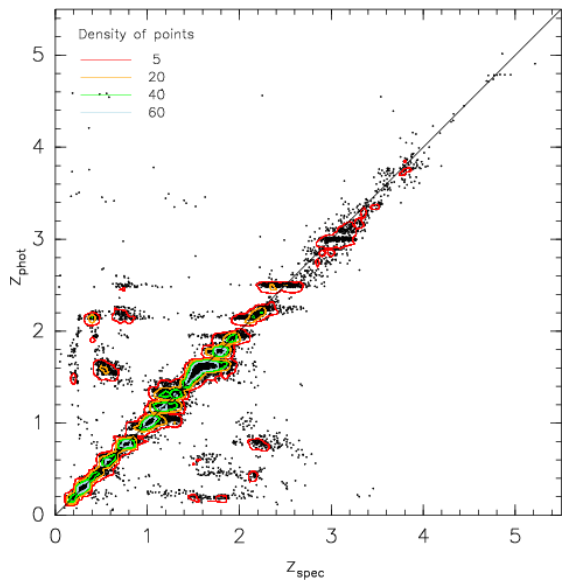}
\caption{As Figure \ref{Fig: SDSS photoz}, but showing the results for the CZR
  photozs. The regions of catastrophic failure are seen, and the overall
  variance is $\variance = 0.265 \pm 0.006$. The values of $\zphot$ resulting 
  from this method are in bins of width 0.05. Here, a uniformly distributed 
  random offset up to $\pm 0.025$ has been added to the values of $\zphot$ for
  clarity.
  \label{Fig: CZR photoz}}
\end{figure}

We have also implemented the methods of W04 on the SDSS DR3, without removing
the reddened quasars (Strand, in preparation). Here we apply that method to the
SDSS DR5 dataset as a direct comparison between the empirical CZR and the IB. We
find that the CZR has slightly narrower dispersion than the IB, with $\dz$
percentages of $63.9 \pm 0.3\%$, $80.2 \pm 0.4\%$ and $85.7 \pm 0.3\%$ within
$\dzrange$. However, as shown in Figure \ref{Fig: CZR photoz}, it still shows
regions of catastrophic failure. The variance is therefore significantly higher,
at $\variance = 0.265 \pm 0.006$. We again plot the run from the ten with the
closest variance to the mean. In this case this was the final run of the ten,
with $\variance = 0.2653$.

Previous results using empirical CZRs show a similar pattern. For example,
Figure 4 of W04 shows regions of quasars at $0 \lesssim \zphot \lesssim 1$ and
$1.5 \lesssim \zphot \lesssim 4.5$ over the spectroscopic redshift range $0
\lesssim \zspec \lesssim 4$. Similar results are seen in
\citet{budavari:qsophotoz}, \citet{richards:qsophotoz}, and
\citet{wu:qsophotoz}.

\subsection{SDSS DR5 + GALEX GR2}

Adding the GALEX data significantly improves the results, as shown in Figures
\ref{Fig: GALEX photoz} and \ref{Fig: GALEX grid}. Here we obtain a variance of
$0.054 \pm 0.005$ for the pseudo-blind test, $70.8 \pm 1.2\%$, $85.8 \pm 1.0\%$
and $90.8 \pm 0.7\%$ of objects within $\dzrange$, $\zvariance = 0.014 \pm
0.002$, and the mean $\dzopz = 0.060 \pm 0.003$.

The number of nearest neighbors and distance weighting are $17 \pm 5$ and $4.4
\pm 0.8$ respectively. A higher distance weighting is expected due to the
greater dimensionality of the training feature space (21 colors instead of 16)
compared to the SDSS dataset.

The exact values of NN and DW that are plotted in Figure \ref{Fig: GALEX photoz}
are chosen in the same manner as for the SDSS, and are ${\mathrm{NN}} = 12$,
${\mathrm{DW}} = 5$ and a random seed of 3. The variance is 0.0521.

To show that the improvement is not simply due to the smaller set of objects
which appear in both surveys (for example, these objects may be brighter quasars
in the SDSS with better photometry), we also applied the SDSS training procedure
to the cross-matched sample. This gives better results than the SDSS sample, but
they are still significantly worse than SDSS+GALEX. The variance is $\variance =
0.090 \pm 0.007$, and the other results are as seen in Table \ref{Table:
  photoz}.

The SDSS results extend deeper than those matched with GALEX, to $z \lesssim 6$
rather than $z \lesssim 3.5$. The lack of quasars in the `redshift desert' at $z
\gtrsim 2.2$ is seen in Figure \ref{Fig: GALEX photoz}, caused by the Lyman
break in the spectrum at a restframe wavelength of $912~{\mathrm{\AA}}$ being
shifted out of the UV.

The CZR results for SDSS+GALEX also improve over those from the full SDSS
dataset. $74.9 \pm 1.4\%$, $86.9 \pm 0.6\%$, and $91.0 \pm 0.8\%$ of the objects
are within $\dzrange$. This is still slightly better than IB for $\dz < 0.1$ and
$\dz < 0.2$, but is the same for $\dz < 0.3$.

\subsection{Genetic Algorithms} \label{Subsec: GA Results}

The application of the genetic algorithms on the SDSS, SDSS+GALEX and
GALEX-SDSS-only datasets converged on the use of approximately half of the
training parameters, but the variance was not significantly different from that
from using the full set of training features. The full sets were therefore used
throughout. The result indicates that there is some redundancy in the training
features, which is expected given that they are measuring the four colors four
different times, just through different apertures.

\section{Discussion} \label{Sec: Discussion}

Although the results here represent an important step in the sense that there
are no regions of catastrophic failure, further improvement is still
possible. In particular: (1) The input object parameter distributions may be
generalized into the form of a PDF for each object, which can be propagated
through the learning process, to make more explicit those objects for which the
redshift is less certain, to take into account the error on each parameter, and
to output a PDF for each object instead of a scalar value. (2) The
no-catastrophics of the instance-based and the lower low-$\dz$ dispersion of the
CZR can be combined into a new learning algorithm. The IB is in fact able to
obtain similar results to the CZR (i.e., an approximately 5\% narrower
dispersion and regions of catastrophic failure instead of a spread of objects),
by using the single nearest neighbor instead of $k$ nearest neighbors. (3) The
addition of other multiwavelength training data, such as infrared data from
UKIDSS \citep{lawrence:ukidss} and Spitzer \citep{werner:spitzer}, can be
included in the training process.

We also obtained quasar photometric redshifts using decision trees, as used in
\citet{ball:dtclassification} for star-galaxy separation. The variances obtained
were generally comparable to, but slightly worse than, those for instance-based,
and are, therefore, not reported here.

\section{Conclusions} \label{Sec: Conclusions}

We apply instance-based machine learning to 55,746 objects spectroscopically
classified as quasars in the Fifth Data Release of the Sloan Digital Sky Survey
(SDSS), and to 7,642 objects cross-matched from this sample to the Second Data
Release of the Galaxy Evolution Explorer legacy data (SDSS+GALEX).

The algorithm is able to assign photometric redshifts to quasars without regions
of catastrophic failure, unlike previously published results. This will enable
samples of quasars to be constructed for cosmological studies with minimal
contamination from objects at severely incorrect redshifts.

We obtain, for the same data, empirical color-redshift relations with full
probability distributions and find that these are similar to previous results in
the literature.

For SDSS, we find a photometric-to-spectroscopic variance of $0.123 \pm 0.002$
for a sample of the data not used in the training. For SDSS+GALEX, this improves
to $0.054 \pm 0.005$. Using purely SDSS on the latter dataset (GALEX-SDSS-only),
the variance is $0.090 \pm 0.007$. Hence the improvement results from the extra
UV information provided by GALEX and not the reduced sample size, better
photometry, or lower redshifts. The percentages of objects within $\dz < 0.1$
are $54.9 \pm 0.7\%$, $70.8 \pm 1.2\%$, and $62.0 \pm 1.4\%$ for SDSS,
SDSS+GALEX, and GALEX-SDSS-only, respectively.

Each set of results represents a realistic standard for application to further
datasets of which the spectra are representative.

\begin{figure}
\figurenum{4}
\plotone{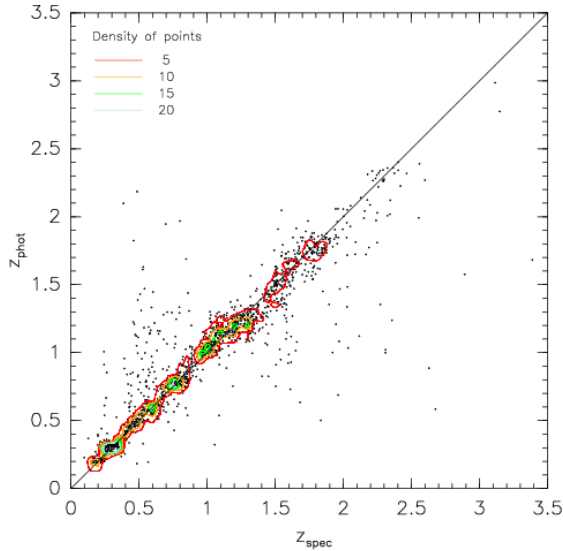}
\caption{As Figure \ref{Fig: SDSS photoz}, but showing the results for 1,528 of
  7,642 quasars present in the SDSS DR5 cross-matched to the GALEX GR2. The
  variance is improved to $\variance = 0.054 \pm 0.005$. \label{Fig: GALEX
    photoz}}
\end{figure}

\begin{figure}
\figurenum{5}
\plotone{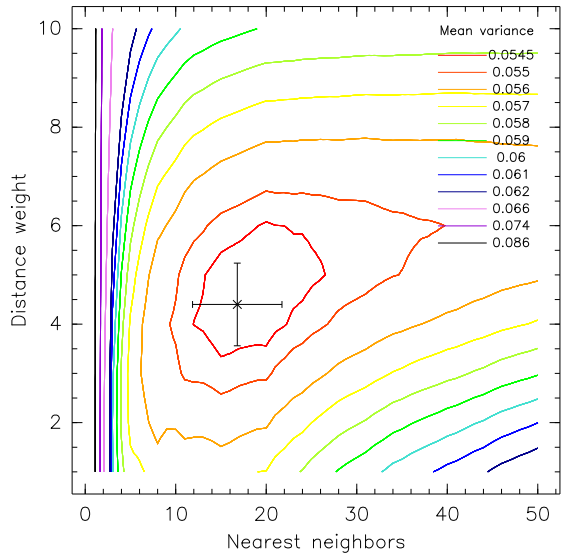}
\caption{As Figure \ref{Fig: SDSS grid}, but for the SDSS+GALEX dataset shown in
  Figure \ref{Fig: GALEX photoz}. \label{Fig: GALEX grid}}
\end{figure}

\begin{deluxetable*}{cccccccc}
\tablenum{1}
\tablewidth{0pt}
\tablecaption{Summary of photometric redshift samples described in this
  paper. \label{Table: photoz}}
\tablehead{\colhead{Dataset} &\colhead{Method} &\colhead{Variance} &\colhead
  {Variance/(1+z)} &\colhead{Mean $\dzopz$} &\colhead{$\% \dz < 0.1$}
  &\colhead{$\% \dz < 0.2$} &\colhead{$\% \dz < 0.3$}}
\startdata
SDSS            &IB  &$0.123 \pm 0.002$ &$0.034 \pm 0.001$ &$0.095 \pm 0.001$
&$54.9 \pm 0.7$ &$73.3 \pm 0.6$ &$80.7 \pm 0.3$\\
SDSS+GALEX      &IB  &$0.054 \pm 0.005$ &$0.014 \pm 0.002$ &$0.060 \pm 0.003$
&$70.8 \pm 1.2$ &$85.8 \pm 1.0$ &$90.8 \pm 0.7$\\
GALEX-SDSS-only &IB  &$0.090 \pm 0.007$ &$0.022 \pm 0.001$ &$0.081 \pm 0.003$
&$62.0 \pm 1.4$ &$78.9 \pm 1.0$ &$85.2 \pm 1.2$\\

SDSS            &CZR &$0.265 \pm 0.006$ &$0.079 \pm 0.003$ &$0.115 \pm 0.002$
&$63.9 \pm 0.3$ &$80.2 \pm 0.4$ &$85.7 \pm 0.3$\\
SDSS+GALEX      &CZR &$0.136 \pm 0.015$ &$0.031 \pm 0.006$ &$0.071 \pm 0.005$
&$74.9 \pm 1.4$ &$86.9 \pm 0.6$ &$91.0 \pm 0.8$\\
GALEX-SDSS-only &CZR &$0.158 \pm 0.013$ &$0.041 \pm 0.004$ &$0.081 \pm 0.004$
&$74.1 \pm 0.8$ &$86.2 \pm 0.7$ &$89.7 \pm 0.6$\\
\enddata
\end{deluxetable*}

\acknowledgments

We thank the referee for a prompt and useful report which improved the paper, 
and Kumara Sastry of the Illinois Genetic Algorithms Laboratory for a 
clarification on our use of specific genetic algorithms.

The authors acknowledge support from NASA through grants NN6066H156 and
05-GALEX05-0036, from Microsoft Research, and from the University of
Illinois. The authors made extensive use of the storage and computing facilities
at the National Center for Supercomputing Applications and thank the technical
staff for their assistance in enabling this work.

Funding for the SDSS and SDSS-II has been provided by the Alfred P. Sloan
Foundation, the Participating Institutions, the National Science Foundation, the
U.S. Department of Energy, the National Aeronautics and Space Administration,
the Japanese Monbukagakusho, the Max Planck Society, and the Higher Education
Funding Council for England. The SDSS Web Site is http://www.sdss.org/.

The SDSS is managed by the Astrophysical Research Consortium for the
Participating Institutions. The Participating Institutions are the American
Museum of Natural History, Astrophysical Institute Potsdam, University of Basel,
Cambridge University, Case Western Reserve University, University of Chicago,
Drexel University, Fermilab, the Institute for Advanced Study, the Japan
Participation Group, Johns Hopkins University, the Joint Institute for Nuclear
Astrophysics, the Kavli Institute for Particle Astrophysics and Cosmology, the
Korean Scientist Group, the Chinese Academy of Sciences (LAMOST), Los Alamos
National Laboratory, the Max-Planck-Institute for Astronomy (MPA), the
Max-Planck-Institute for Astrophysics (MPIA), New Mexico State University, Ohio
State University, University of Pittsburgh, University of Portsmouth, Princeton
University, the United States Naval Observatory, and the University of
Washington. 


Based on observations made with the NASA Galaxy Evolution Explorer. GALEX is
operated for NASA by the California Institute of Technology under NASA contract
NAS5-98034. 

Data To Knowledge (D2K) software, D2K modules, and/or D2K itineraries, used by
us, were developed at the National Center for Supercomputing Applications (NCSA)
at the University of Illinois at Urbana-Champaign.

This research has made use of NASA's Astrophysics Data System.


\begin{thebibliography}{60}
\expandafter\ifx\csname natexlab\endcsname\relax\def\natexlab#1{#1}\fi

\bibitem[{{Abazajian} {et~al.}(2004)}]{abazajian:dr2}
{Abazajian}, K. {et~al.} 2004, \aj, 128, 502

\bibitem[{{Aha} {et~al.}(1991){Aha}, Kibler, \& Albert}]{aha:ib}
{Aha}, D.~W., Kibler, D., \& Albert, M.~K. 1991, Machine Learning, 6, 37

\bibitem[{{Babbedge} {et~al.}(2004)}]{babbedge:impz}
{Babbedge}, T.~S.~R. {et~al.} 2004, \mnras, 353, 279

\bibitem[{{Ball} {et~al.}(2006){Ball}, {Brunner}, {Myers}, \&
  {Tcheng}}]{ball:dtclassification}
{Ball}, N.~M., {Brunner}, R.~J., {Myers}, A.~D., \& {Tcheng}, D. 2006, \apj,
  650, 497

\bibitem[{{Ball} {et~al.}(2004){Ball}, {Loveday}, {Fukugita}, {Nakamura},
  {Okamura}, {Brinkmann}, \& {Brunner}}]{ball:ann}
{Ball}, N.~M., {Loveday}, J., {Fukugita}, M., {Nakamura}, O., {Okamura}, S.,
  {Brinkmann}, J., \& {Brunner}, R.~J. 2004, \mnras, 348, 1038

\bibitem[{{Baum}(1962)}]{baum:photoz}
{Baum}, W.~A. 1962, in IAU Symp. 15: Problems of Extra-Galactic Research, 390

\bibitem[{{Beckwith} {et~al.}(2006)}]{beckwith:hudf}
{Beckwith}, S.~V.~W. {et~al.} 2006, \aj, 132, 1729

\bibitem[{{Ben{\'{\i}}tez}(2000)}]{benitez:photoz}
{Ben{\'{\i}}tez}, N. 2000, \apj, 536, 571

\bibitem[{{Brunner} {et~al.}(1997){Brunner}, {Connolly}, {Szalay}, \&
  {Bershady}}]{brunner:photoz}
{Brunner}, R.~J., {Connolly}, A.~J., {Szalay}, A.~S., \& {Bershady}, M.~A.
  1997, \apjl, 482, L21

\bibitem[{{Brunner} {et~al.}(2000){Brunner}, {Szalay}, \&
  {Connolly}}]{brunner:angcf}
{Brunner}, R.~J., {Szalay}, A.~S., \& {Connolly}, A.~J. 2000, \apj, 541, 527

\bibitem[{{Budav{\'a}ri} {et~al.}(2001)}]{budavari:qsophotoz}
{Budav{\'a}ri}, T. {et~al.} 2001, \aj, 122, 1163

\bibitem[{{Cannon} {et~al.}(2006)}]{cannon:2slaqlrg}
{Cannon}, R. {et~al.} 2006, \mnras, 372, 425

\bibitem[{{Cardelli} {et~al.}(1989){Cardelli}, {Clayton}, \&
  {Mathis}}]{cardelli:extinction}
{Cardelli}, J.~A., {Clayton}, G.~C., \& {Mathis}, J.~S. 1989, \apj, 345, 245

\bibitem[{{Coe} {et~al.}(2006){Coe}, {Ben{\'{\i}}tez}, {S{\'a}nchez}, {Jee},
  {Bouwens}, \& {Ford}}]{coe:photoz}
{Coe}, D., {Ben{\'{\i}}tez}, N., {S{\'a}nchez}, S.~F., {Jee}, M., {Bouwens},
  R., \& {Ford}, H. 2006, \aj, 132, 926

\bibitem[{{Collister} {et~al.}(2007)}]{collister:megazlrg}
{Collister}, A. {et~al.} 2007, \mnras, 375, 68

\bibitem[{{Collister} \& {Lahav}(2004)}]{collister:annz}
{Collister}, A.~A. \& {Lahav}, O. 2004, \pasp, 116, 345

\bibitem[{{Connolly} {et~al.}(1998){Connolly}, {Szalay}, \&
  {Brunner}}]{connolly:angcf}
{Connolly}, A.~J., {Szalay}, A.~S., \& {Brunner}, R.~J. 1998, \apjl, 499, L125

\bibitem[{{Cover} \& {Hart}(1967)}]{cover:nn}
{Cover}, T.~M. \& {Hart}, P.~E. 1967, IEEE Transactions on Information Theory,
  13, 21

\bibitem[{{Croom} {et~al.}(2004){Croom}, {Smith}, {Boyle}, {Shanks}, {Miller},
  {Outram}, \& {Loaring}}]{croom:qsolf}
{Croom}, S.~M., {Smith}, R.~J., {Boyle}, B.~J., {Shanks}, T., {Miller}, L.,
  {Outram}, P.~J., \& {Loaring}, N.~S. 2004, \mnras, 349, 1397

\bibitem[{{Csabai} {et~al.}(2003)}]{csabai:edrphotoz}
{Csabai}, I. {et~al.} 2003, \aj, 125, 580

\bibitem[{{Eisenstein} {et~al.}(2001)}]{eisenstein:lrgsample}
{Eisenstein}, D.~J. {et~al.} 2001, \aj, 122, 2267

\bibitem[{{Firth} {et~al.}(2003){Firth}, {Lahav}, \&
  {Somerville}}]{firth:annphotoz}
{Firth}, A.~E., {Lahav}, O., \& {Somerville}, R.~S. 2003, \mnras, 339, 1195

\bibitem[{{Goldberg}(1989)}]{goldberg:genetic}
{Goldberg}, D.~E. 1989, {Genetic Algorithms in Search, Optimization, and
  Machine Learning} (Reading, MA: Addison-Wesley)

\bibitem[{Goldberg(2002)}]{goldberg:design}
Goldberg, D.~E. 2002, Design of innovation: {L}essons from and for competent
  genetic algorithms (Boston, MA: Kluwer Academic Publishers)

\bibitem[{{Gwyn} \& {Hartwick}(1996)}]{gwyn:photoz}
{Gwyn}, S.~D.~J. \& {Hartwick}, F.~D.~A. 1996, \apjl, 468, L77

\bibitem[{{Hastie} {et~al.}(2001){Hastie}, {Tibshirani}, \&
  {Friedman}}]{hastie:learning}
{Hastie}, T., {Tibshirani}, R., \& {Friedman}, J. 2001, {The Elements of
  Statistical Learning} (Springer)

\bibitem[{{Haupt} \& {Haupt}(1998)}]{haupt:genetic}
{Haupt}, R.~L. \& {Haupt}, S.~E. 1998, {Practical Genetic Algorithms} (New
  York: Wiley Inter-Science)

\bibitem[{{Hogg} {et~al.}(1998)}]{hogg:photoz}
{Hogg}, D.~W. {et~al.} 1998, \aj, 115, 1418

\bibitem[{{Holland}(1975)}]{holland:genetic}
{Holland}, J.~H. 1975, {Adaptation in Natural and Artificial Systems: An
  Introductory Analysis with Applications to Biology, Control and Artificial
  Intelligence} (Ann Arbor, MI: The University of Michigan Press)

\bibitem[{{Koo}(1985)}]{koo:photoz}
{Koo}, D.~C. 1985, \aj, 90, 418

\bibitem[{{Lanzetta} {et~al.}(1996){Lanzetta}, {Yahil}, \&
  {Fernandez-Soto}}]{lanzetta:photoz}
{Lanzetta}, K.~M., {Yahil}, A., \& {Fernandez-Soto}, A. 1996, \nat, 381, 759

\bibitem[{{Lawrence} {et~al.}(2006)}]{lawrence:ukidss}
{Lawrence}, A. {et~al.} 2006, preprint, astro-ph/0604426

\bibitem[{{Loh} \& {Spillar}(1986)}]{loh:photoz}
{Loh}, E.~D. \& {Spillar}, E.~J. 1986, \apj, 303, 154

\bibitem[{{Martin} {et~al.}(2005)}]{martin:galex}
{Martin}, D.~C. {et~al.} 2005, \apjl, 619, L1

\bibitem[{{Mobasher} {et~al.}(1996){Mobasher}, {Rowan-Robinson}, {Georgakakis},
  \& {Eaton}}]{mobasher:photoz}
{Mobasher}, B., {Rowan-Robinson}, M., {Georgakakis}, A., \& {Eaton}, N. 1996,
  \mnras, 282, L7

\bibitem[{{Myers} {et~al.}(2007{\natexlab{a}}){Myers}, {Brunner}, {Nichol},
  {Richards}, {Schneider}, \& {Bahcall}}]{myers:nbckdeall}
{Myers}, A.~D., {Brunner}, R.~J., {Nichol}, R.~C., {Richards}, G.~T.,
  {Schneider}, D.~P., \& {Bahcall}, N.~A. 2007{\natexlab{a}}, ApJ, in press
  (astro-ph/0612190)

\bibitem[{{Myers} {et~al.}(2007{\natexlab{b}}){Myers}, {Brunner}, {Richards},
  {Nichol}, {Schneider}, \& {Bahcall}}]{myers:nbckdesmall}
{Myers}, A.~D., {Brunner}, R.~J., {Richards}, G.~T., {Nichol}, R.~C.,
  {Schneider}, D.~P., \& {Bahcall}, N.~A. 2007{\natexlab{b}}, ApJ, in press,
  (astro-ph/0612191)

\bibitem[{{Myers} {et~al.}(2006)}]{myers:qsoevoln}
{Myers}, A.~D. {et~al.} 2006, \apj, 638, 622

\bibitem[{{Padmanabhan} {et~al.}(2005)}]{padmanabhan:photoz}
{Padmanabhan}, N. {et~al.} 2005, \mnras, 359, 237

\bibitem[{{Richards} {et~al.}(2001)}]{richards:qsophotoz}
{Richards}, G.~T. {et~al.} 2001, \aj, 122, 1151

\bibitem[{{Sawicki} {et~al.}(1997){Sawicki}, {Lin}, \& {Yee}}]{sawicki:photoz}
{Sawicki}, M.~J., {Lin}, H., \& {Yee}, H.~K.~C. 1997, \aj, 113, 1

\bibitem[{{Schlegel} {et~al.}(1998){Schlegel}, {Finkbeiner}, \&
  {Davis}}]{schlegel:dustmaps}
{Schlegel}, D.~J., {Finkbeiner}, D.~P., \& {Davis}, M. 1998, \apj, 500, 525

\bibitem[{{Skrutskie} {et~al.}(2006)}]{skrutskie:2mass}
{Skrutskie}, M.~F. {et~al.} 2006, \aj, 131, 1163

\bibitem[{{Stoughton} {et~al.}(2002)}]{stoughton:edr}
{Stoughton}, C. {et~al.} 2002, \aj, 123, 485

\bibitem[{Tagliaferri {et~al.}(2002)Tagliaferri, Longo, Andreon, Capozziello,
  Donalek, \& Giordanoet}]{tagliaferri:nnphotoz}
Tagliaferri, R., Longo, G., Andreon, S., Capozziello, S., Donalek, C., \&
  Giordanoet, G. 2002, preprint, astro-ph/0203445

\bibitem[{{Vanden Berk} {et~al.}(2005)}]{vandenberk:empirical}
{Vanden Berk}, D.~E. {et~al.} 2005, \aj, 129, 2047

\bibitem[{{Vanzella} {et~al.}(2004)}]{vanzella:hdfannphotoz}
{Vanzella}, E. {et~al.} 2004, \aap, 423, 761

\bibitem[{{Wadadekar}(2005)}]{wadadekar:photoz}
{Wadadekar}, Y. 2005, \pasp, 117, 79

\bibitem[{{Wang} {et~al.}(1998){Wang}, {Bahcall}, \& {Turner}}]{wang:photoz}
{Wang}, Y., {Bahcall}, N., \& {Turner}, E.~L. 1998, \aj, 116, 2081

\bibitem[{{Way} \& {Srivastava}(2006)}]{way:photoz}
{Way}, M.~J. \& {Srivastava}, A.~N. 2006, \apj, 647, 102

\bibitem[{{Weinstein} {et~al.}(2004)}]{weinstein:sdssqsophotoz}
{Weinstein}, M.~A. {et~al.} 2004, \apjs, 155, 243

\bibitem[{{Welge} {et~al.}(1999){Welge}, {Hsu}, {Auvil}, {Redman}, \&
  {Tcheng}}]{welge:d2k}
{Welge}, M., {Hsu}, W.~H., {Auvil}, L.~S., {Redman}, T.~M., \& {Tcheng}, D.
  1999, in 12th National Conference on High Performance Networking and
  Computing (SC99)

\bibitem[{{Werner} {et~al.}(2004)}]{werner:spitzer}
{Werner}, M.~W. {et~al.} 2004, \apjs, 154, 1

\bibitem[{{Williams} {et~al.}(1996)}]{williams:hdfn}
{Williams}, R.~E. {et~al.} 1996, \aj, 112, 1335

\bibitem[{{Williams} {et~al.}(2000)}]{williams:hdfs}
---. 2000, \aj, 120, 2735

\bibitem[{{Witten} \& {Frank}(2000)}]{witten:datamining}
{Witten}, I.~H. \& {Frank}, E. 2000, {Data Mining} (Morgan Kaufmann)

\bibitem[{{Wolf} {et~al.}(2003){Wolf}, {Wisotzki}, {Borch}, {Dye},
  {Kleinheinrich}, \& {Meisenheimer}}]{wolf:qsophotoz}
{Wolf}, C., {Wisotzki}, L., {Borch}, A., {Dye}, S., {Kleinheinrich}, M., \&
  {Meisenheimer}, K. 2003, \aap, 408, 499

\bibitem[{{Wolf} {et~al.}(2004)}]{wolf:combo17}
{Wolf}, C. {et~al.} 2004, \aap, 421, 913

\bibitem[{{Wu} {et~al.}(2004){Wu}, {Zhang}, \& {Zhou}}]{wu:qsophotoz}
{Wu}, X.-B., {Zhang}, W., \& {Zhou}, X. 2004, Chinese Journal of Astronomy and
  Astrophysics, 4, 17

\bibitem[{{York} {et~al.}(2000)}]{york:sdss}
{York}, D.~G. {et~al.} 2000, \aj, 120, 1579

\end{thebibliography}

\end{document}